**A Model for Optimizing the Health and Economic Impacts of Covid-19 under Social Distancing Measures; A Study for the Number of Passengers and their Seating Arrangements in Aircrafts**

Elaheh Ghorbani[1,2*], Hamid Molavian[2,3*], Fred Barez[1*]

[1]Department of Aviation and Technology Study, San Jose State University, San Jose, California, USA 95192

[2]Magic Distance, San Jose, California, USA 95126

[3]Data Science Immersive Program, Galvanize Inc., San Francisco, California, USA 94105

## Abstract

Covid-19 has had a disastrous economic impact on countries and industries as countries have gone through the lockdown process to reduce the health impact of Covid-19. As countries have started lifting Covid-19 related restrictions, businesses have been allowed to again have on-site customers. However, just a limited number of people are being allowed on-site as long as social distancing measures are being followed. This has resulted in heavy burdens on businesses as their number of customers have decreased substantially. In this study, we propose a model to minimize the economic impact of Covid-19 for businesses that have implemented social distancing measures, as well as to minimize the health impact of Covid-19 for their customers and employees. We introduce the quantity Spread in which minimizing Spread gives the optimum number and arrangement of people at a given site while applying social distancing measures. We apply our model to a real-world scenario and optimize the number of passengers and their arrangements under a social distancing measure for two different popular aircraft seat layouts using the Annealing Monte Carlo technique. We obtain the optimal numbers and optimal arrangements of passengers considering both family groups and individual passengers for the social distancing measure. The obtained optimal arrangements of passengers show complex patterns with groups and individual passengers mixed in complex and non-trivial ways. This demonstrates the necessity of using our model or its variants to find these optimal arrangements. In addition, we show that any other arrangements of passengers with the same number of passengers is a suboptimal arrangement with higher health risks as a result of less distance between passengers. Our model could be implemented for other social situations such as sports events, theaters, medical centers, etc.

## Introduction

Simultaneously minimizing the health and economic impacts of Covid-19 is the biggest challenge for countries and industries as the lockdown has become the first line of defense against Covid-19 and many countries have gone through the lockdown phase [1-4]. As the economic damages of Covid-19 have become unbearable as a result of lockdown, countries have begun the reopening phases and have allowed their businesses to reopen to avoid the catastrophic economic effects of Covid-19 [5,6]. In the reopening phases many businesses and countries have implemented restrictive measures including social distancing measures as it has been strongly argued to be the most effective non-pharmacological way to reduce the health impact of Covid-19 [7-10]. Complementary methods such as washing hands and using masks have been also recommended strongly or enforced by businesses and governments [11,12]. Also, many studies have been done to either introduce strategies to optimize the current restrictive methods or to investigate the health and economic impacts of current restrictions on people [13-28]. Additionally, many new products such as physical barriers, covers and protocols have been introduced to the market to help different businesses implement social distancing measures more rigorously and reduce the health risk of Covid-19 for their on-site employees and customers. However, the main way of applying the social distancing measures has been through leaving empty spaces between people using signages on floors and seats. These signages have been installed in simple ways such as leaving empty rows or columns of seats between people to create social distancing (usually 6 feet). To the best of our knowledge, it seems that none of the businesses and industries have used modeling and optimization methods to maximize the number of people on-site and their distances, under a social distancing constraint, to decrease the health and economic impacts of Covid-19. Our modeling and optimization methods could be applied to a variety of settings, such as classrooms, restaurants, theaters, medical centers and etc.

In this paper we report the results of optimizing the number of passengers and their arrangements under a social distancing measure for the airline industry by introducing a model. In our model we introduce the quantity Spread and by applying simulated Annealing Monte Carlo we minimize Spread to obtain the optimum number and arrangement of passengers for two types of single and twin aisle aircraft layouts under a social distancing measure. We show that our obtained optimal arrangements have higher distance between passengers as compared to any suboptimal arrangements with the same number of passengers. Our results suggest that it is necessary to use modeling methods to minimize the health and economic impacts of Covid-19 under social distancing measures.

## Results

### Model

We introduce our model based on the observed facts about spreading the SARS-CoV-2 virus as well as the social distancing measures that have been suggested by health experts and enforced or recommended by businesses and governments. Social distancing is the physical distance between two people and has been recommended to be 6 feet. In this model, family groups, or people who live together, do not need to follow the social distancing measure. However, individuals and groups that do not live together do need to follow the social distancing measure. Based on these facts, we introduce the below model with a quantity that we name Spread, and we minimize Spread to obtain the optimum number of passengers and their arrangements. We assume that we have a two-dimensional problem with all people being assigned a position in an xy plane. Based on these facts we define the Spread for person i ($spread_i$) as follows:

$$spread_i = \begin{cases} \sum_{i \neq j}(r_{ij} - 1)^2 & if\ r_{ij} \leq 1,\ i\ and\ j\ are\ in\ the\ same\ group \\ \sum_{i \neq j}(r_{ij} - 2)^2 + (y_i - y_j)^2 & if\ r_{ij} > 1,\ i\ and\ j\ are\ in\ the\ same\ group \\ \sum_{i \neq j} 2(sd - 1)^2 & if\ r_{ij} < sd,\ i\ and\ j\ are\ not\ in\ the\ same\ group \\ \sum_{i \neq j} \frac{\delta}{r_{ij}^2} & if\ r_{ij} \geq sd,\ i\ and\ j\ are\ not\ in\ the\ same\ group \end{cases} \quad (1)$$

where $r_{ij}$ and $y_i - y_j$ are, respectively, the distance and the distance in the y direction between individuals i and j, $sd$ is the minimum social distancing that should be followed by individuals, $\delta$ = 1 and its unit is $[r_{ij}^4]$ .

Eq. 1 has four terms. The first and second terms keep the family members close to each other. Term $(y_i - y_j)^2$ in the second equation dictates that the members of a family group sit close to each other on the same row and term $(r - 2)^2$ dictates that the members of the same family group stay as close as possible to each other. We should note that other terms such as $(r - n)^2$, where $n$ = 3, 4, …, might be added to the second term if it is desirable to keep big family groups close to each other. The third equation in our model is the penalty for not following the social distancing measure for people that do not belong to the same family group. The fourth equation keeps people who are not in the same family group as far as possible from each other and it acts as a repulsion force. We define the total Spread as sum of the individual Spreads as follows:

$$S_{tot} = \sum_i spread_i \quad (2)$$

Where sum is over all on-site people and our goal is to minimize $S_{tot}$ to obtain the maximum distance between non-group individuals who follow the social distancing measure. Since the number of passengers under the social distancing is less than the number of seats, the number of possible arrangements of passengers is $\binom{n}{m}$, where $n$ and $m$ are the number of seats and the number of passengers, respectively. The number of possible arrangements can grow exponentially; for instance, for a half-filled aircraft with 96 seats, the number of possible arrangements is $\sim 2^{96} \sim 8 \times 10^{28}$. Hence, we use the simulated Annealing Monte Carlo [29, 30] approaches to minimize $S_{tot}$ and obtain the optimal arrangements and number of passengers.

## Seat Layouts and Parameters

In this work we consider two popular seat layouts in aircrafts, namely, single (Fig. 1a) and twin (Fig. 1b) aisles cabin layouts for the economy class. The same model could be applied to the business class. The single aisle layout has 16 rows, with 6 seats in each row, separated evenly by an aisle, making a total of 96 seats (Fig 1a). The twin aisle layout has 17 rows, with 7 seats in each row, 2 seats on either side and 3 seats in the center, making a total of 119 seats (Fig. 1b). These single and twin aisle layouts correspond to popular seat layouts in Boeing 737-800 and Boeing 767, respectively. There are, however, variations of these layouts. In both of these layouts, we consider that the distance between two adjacent seats is one unit, the distance between a seat and its front and back seat is 1.8 units, and the width of the aisle is one unit as well (Fig. 1).

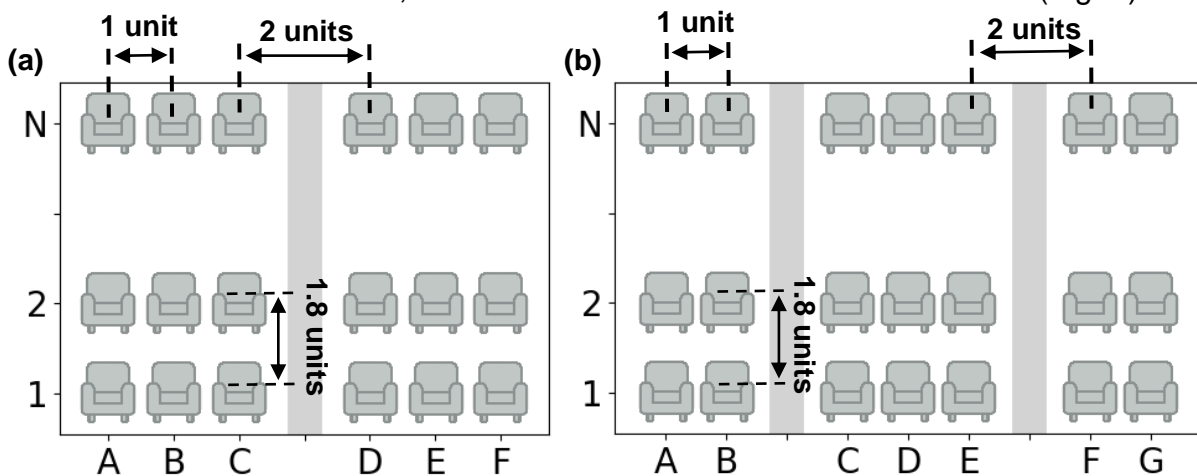

Fig. 1. The arrangement of seats and the unit distance we used in our models for single aisle (a) and twin aisle (b) seat layouts. We assume that the distance between adjacent seats in one row is one unit and the distance of a front and back seat in the same column is 1.8 units.

We use *Spread,* the quantity that we introduced in the Model section and minimize Spread to obtain the arrangement that maximizes the distance between passengers. As we explained in the Model section, we use the simulated Annealing Monte Carlo (AMC) to find the minimum Spread. In the Method section we discuss in detail the Algorithm that we use to minimize Spread.

## Passengers Arrangements for Single Aisle Layout

We begin with cases in which the passengers are a combination of individual passengers and family groups with different number of members and the social distancing guideline of 2.25 units, where unit is defined in the Seat Layouts and Parameters section. Throughout the rest of the paper and in the figures, we show the individual passengers with white masks and family groups with the same color masks. We consider three different cases of passengers as follows: (a) one four-member, two three-member, and four two-member family groups (b) three four-member, one three-member, and five two-member family groups and, (c) four three-member and five two-member family groups in a single layout aircraft. For each of these cases, we maximized the number of individual passengers that can be fitted to the seats under the social distancing measure (2.25 units). In Fig. 2, we show the optimal arrangements of passengers for the mentioned three cases under the social distancing measure.

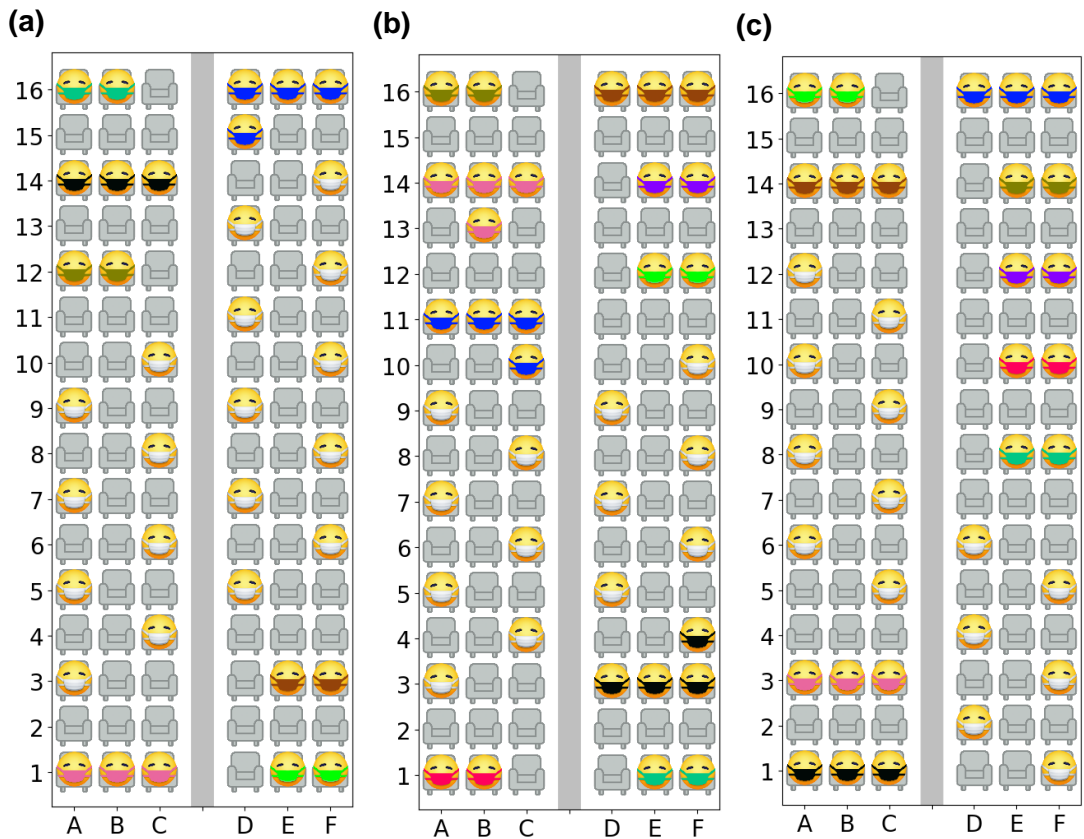

Fig. 2. The optimal arrangements of passengers for (a) one four-member, two three-member, four two-member family groups and eighteen individual passengers (b) three four-member, one three-member, five two-member family groups and thirteen individual passengers (c) four three-member and five two-member family groups and fourteen individual passengers. The arrangements of family groups and individual passengers are determined by the model such that it maximizes the distance between non family group members. Therefore, the location of family groups changes based on the number of family groups and individual passengers.

Our results show that the maximum number of individual passengers that can be fitted to the seats for cases (a), (b) and (c) are eighteen, thirteen, and fourteen, respectively. We observe that, in all these arrangements, family group members are correctly located close to each other and in the same row, when possible, and individual passengers and groups members who do not belong to the same family group follow the social distancing measure. Moreover, when possible, our model keeps the family members as close as possible to each other. This can be seen clearly in the family group with pink masks in Fig. 2b in which the fourth group member is seated in the middle seat close to all the three family members in the back seats. When it is not possible to keep the family members as close as possible, our model correctly shifts a family member to another seat (groups with blue masks in Figures. 2a and 2b) to maximize the number of passengers. Also, in these arrangements, family groups and individual passengers are spread without any specific pattern as the model maximizes the distance between passengers that do not belong to a family group. For example, In Fig. 2a, two of the two-member family groups in rows 12 and 16 are separated by three rows and two of them with one row (seated in rows 1 and 3) as opposed to Fig. 2c where all two-member family groups are separated, one after the other, by one row (rows 8, 10, 12, 14), with one two-member family group located in the top left seats (row 16). We can observe the same behaviors for three-member and four-member family groups in Figures 2a, 2b and 2c. We should also note that there are a large number of arrangements in which only two passengers don't follow the social distancing guidelines (not shown here). In some of these arrangements, all the passengers need to be rearranged to find the optimal arrangement in which passengers follow the social distancing guidelines. Our model, along with simulated Annealing Monte Carlo, is clearly able to distinguish these arrangements from the optimal arrangement that follows the social distancing measures. Moreover, in many cases, there are arrangements that have the same number of passengers with the same number of groups and individual passengers as the optimal arrangement, with passengers following the social distancing measures as well, however, the distances between passengers are not maximized. We refer to these arrangements as suboptimal arrangements, and in the next section we discuss them in detail. Our model is clearly able to distinguish between these suboptimal arrangements and the optimal arrangement and find the optimal arrangement.

## Optimal vs Suboptimal Arrangements

We have done extensive simulations for different cases of groups and individual passengers to find the optimal arrangements and number of passengers and then compare them with suboptimal arrangements. In Fig. 3, we show our results for the average total distance between non family passengers per passenger for these cases for a single aisle layout. These results demonstrate that the average total distance could be increased by more than 10% in the optimal arrangements as compared to suboptimal arrangements. We also find that some of the arrangements do not have suboptimal arrangements or, in some cases, suboptimal arrangements have an average total distance which is very close to the optimal arrangements. We should note that finding the optimal number and arrangements of passengers is almost impossible by trial and error.

To investigate the optimal and suboptimal arrangements in more detail, we discuss one case with two four-member, three three-member, and five two-member family groups for single and twin aisles layouts. We begin with the single aisle layout and find that the maximum number of individual passengers that can be fitted with these groups is eleven passengers. In Fig. 4a, we demonstrate the optimal arrangement of passengers for the groups along with eleven individual passengers. In Fig. 4b, we show a suboptimal arrangement, which, as we defined, has the same numbers of groups and individual passengers as the optimal arrangement, and passengers follow the social distancing measure. To compare the difference between these two arrangements, in Figures 5a and 5b, we, respectively, plot the distance of passengers from other passengers for the optimal and suboptimal arrangements corresponding to the seat arrangement in Figures 4a and 4b. For clarity in the colorbar, we not only show the distance but also the percentage that each color occupies in the plots. From these plots, we see that the red area in the optimal arrangement occupies less area than in the suboptimal arrangement. This demonstrates that, in the optimal arrangement, the overall distance between passengers is higher although the number of passengers in both arrangements is the same, and, in both arrangements, the social distancing measure is followed.

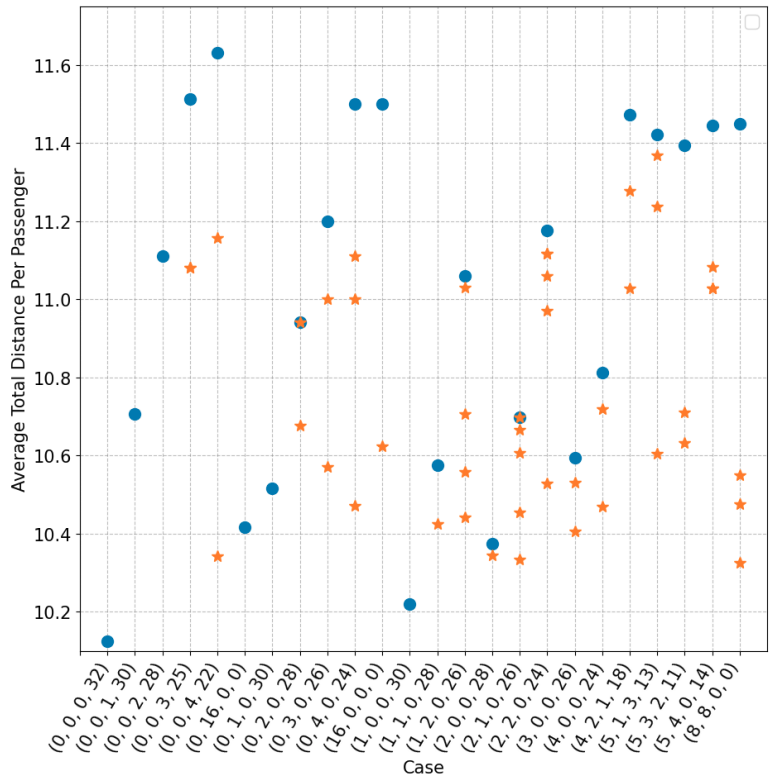

Fig. 3. The average total distance between non family passengers per passenger for some optimal arrangements versus suboptimal arrangements. The optimal arrangements are shown by full blue circles and suboptimal arrangements by orange stars. There might be more than one suboptimal arrangement for each optimal arrangement. The number of groups and individuals are shown in x-axis as tuples with the following structure (Number of four-member family group, Number of three-member family group, Number of two-member family group, Number of individual passengers). Some of the suboptimal arrangements have a closer average total distance as compared to others, depending on the number of groups, number of members in groups, and number of individual passengers. Also, the number of suboptimal arrangements

changes from case to case, depending on the number of groups, number of members in groups and number of individual passengers. In some cases, suboptimal arrangements are very close to optimal arrangements.

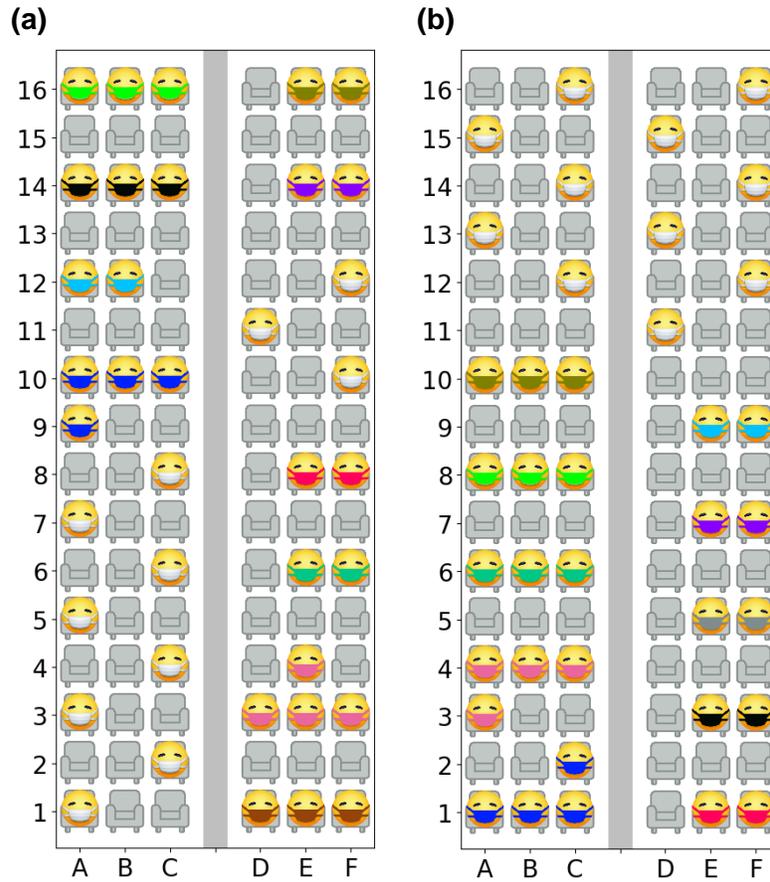

Fig 4. The arrangement of passengers with two four-member, three three-member and five two-member family groups and eleven individual passengers in the optimal (a) and suboptimal (b) arrangements. In both arrangements the social distancing measure is followed by passengers.

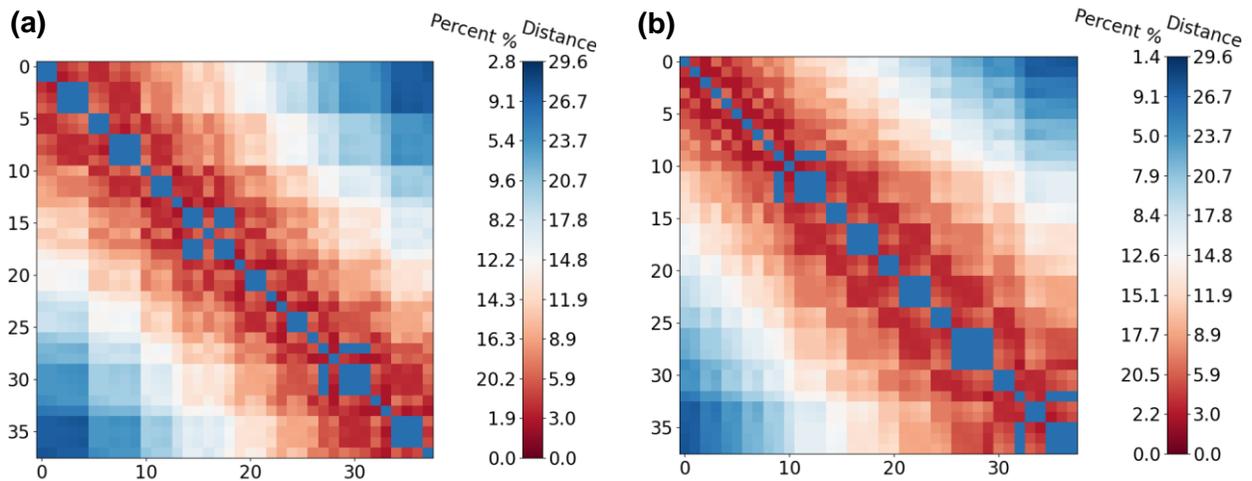

Fig. 5. The counterplot of distance of passengers from other passengers for the optimal (a) and suboptimal (b) arrangements corresponding to Figures 4a and 4b, respectively. From these plots

we see that the suboptimal arrangement has a higher red area than blue, which means that the distance of passengers is higher in the optimal arrangement than the suboptimal arrangement. In the colorbar we show the percentage that each color occupies in the plots.

Since passengers who are close to each other have higher health impacts on each other we define quantity $d_{eff}^j = \frac{1}{\sum_i 1/r_{ij}}$, to measure the health impact of close non family passengers to a passenger (passenger $j$) with the assumption that this impact decreases with the inverse of distance between passengers ($1/r_{ij}$). In this equation $r_{ij}$ is the distance between passenger $i$ and $j$ and the sum is over all the non family groups that are in the range of 4.3 units distance from passenger $j$. $d_{eff}^j$ has the dimension of distance and we refer to it as effective distance throughout the rest of this paper, and it decreases as the distance between passenger $j$ and other non family passengers who are in their vicinity of 4.3 units decreases. We use 4.3 units to only consider the health effects up to three possible nearest passengers. In Figures 6a and 6b we respectively plot effective distance for the two arrangements corresponding to Figures 4a and 4b and in the colorbar we not only show the color related to each effective distance but also the percentage that each color occupies in the arrangement.

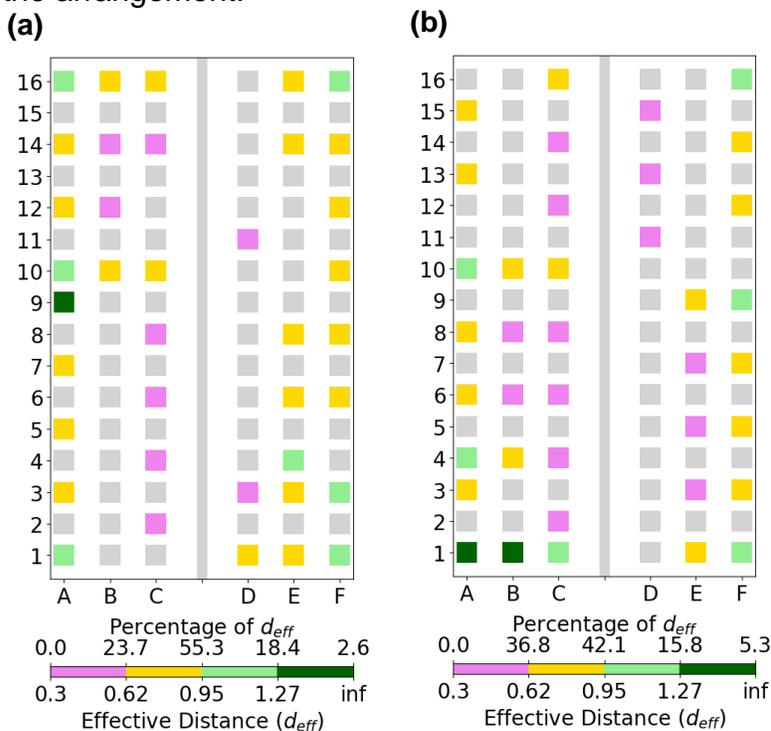

Fig. 6. The effective distance of each passenger from other passengers ($d_{eff}^j$) for the optimal (a) and suboptimal (b) arrangements corresponding to the passengers' arrangements in Figures 4a and 4b, respectively. For clarity, we show the percentage that each color occupies in each plot in the corresponding colorbars. Although a few passengers in the suboptimal arrangement have higher effective distance as compared to others, this is at the cost of all other passengers sitting closer to each other. Overall, in the optimal arrangement, passengers have higher effective distance from each other as compared to the suboptimal arrangement.

Interestingly, from the first look it seems that in the suboptimal arrangement the dark green seats, which corresponds to highest effective distance, occupy 5.3% as compared to the optimal arrangement in which they occupy 2.6%. However, looking more closely we observe that in the optimal case the percentage of passengers that have the closest effective distance (violet color) is 23.7% as compared to 36.8% for the suboptimal arrangement. In addition, gold and light green seats in the optimal case occupy 55.3% and 18.4%, respectively, as compared to the suboptimal arrangement in which gold and light green seats occupy 42.1% and 15.8%. Therefore, in the optimal arrangement, the overall effective distance between passengers is maximized as opposed to the suboptimal arrangement in which a few passengers are seated farther away from others at the cost of making all other passengers seated closer to each other. We should add that seats with higher effective distances can be assigned to passengers that are more vulnerable to Covid-19.

## Passengers Arrangements for Twin Aisle Layout

For the second case, we consider a twin aisle layout with the same number of family groups as the single aisle layout and find the optimum number of individual passengers that can be fitted to the seats. Our results show that the optimal number of individual passengers that can be fitted to the seats with the groups is nineteen. In Fig. 7a we demonstrate the optimal arrangement and, similarly to the single aisle case, we show the results of a suboptimal arrangement with the same number of groups and individual passengers (Fig. 7b). To compare these two arrangements, in Figures 8a and 8b we show the distance of each passenger from other passengers corresponding to the arrangements in Figures 7a and 7b. The bigger red area for the suboptimal arrangements clearly shows that in the optimal arrangement the distance between passengers is higher than in the suboptimal arrangement. In Figures 9a and 9b, we, respectively, plot the effective distance (defined in the previous section) of each passenger from other passengers corresponding to the passenger arrangements in Figures 7a and 7b. From these two plots we clearly see that, overall, the passengers in the optimal arrangements have a higher effective distance than suboptimal arrangement. This could be seen more clearly from the percentage of area that each color occupies (shown in the colorbar). Unlike the single aisle case, in this case, we observe that in the optimal arrangement the dark green area occupies 8.7% of the seats as compared to 4.3% for the suboptimal arrangement. This shows that, in some cases, our model not only maximizes the overall distance between passengers but also creates more seats that have the highest distance from other seats as compared to suboptimal cases. We should also add that, in the optimal arrangement, the light green, gold and violet seats occupy 13.0%, 32.6%, and 45.7% of seats, respectively, as compared to the suboptimal case, in which light green, gold, and violet occupy 13%, 19.6%, and 63.0% of seats, respectively.

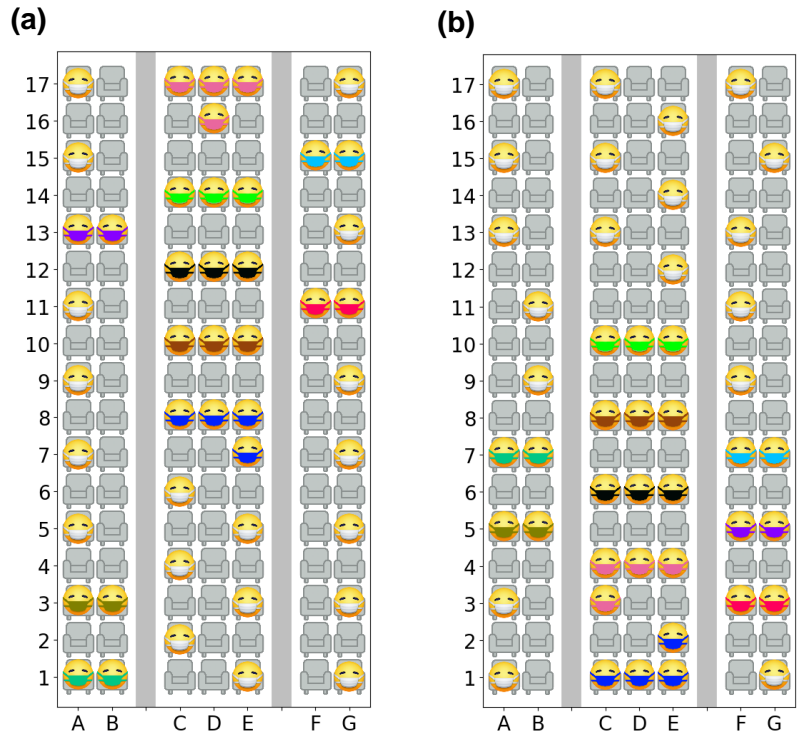

Fig. 7. The arrangement of passengers with two four-member, three three-member, and five two-member family groups and nineteen individual passengers in the optimal (a) and the suboptimal (b) arrangement. In both arrangements, the social distancing measure is followed by passengers.

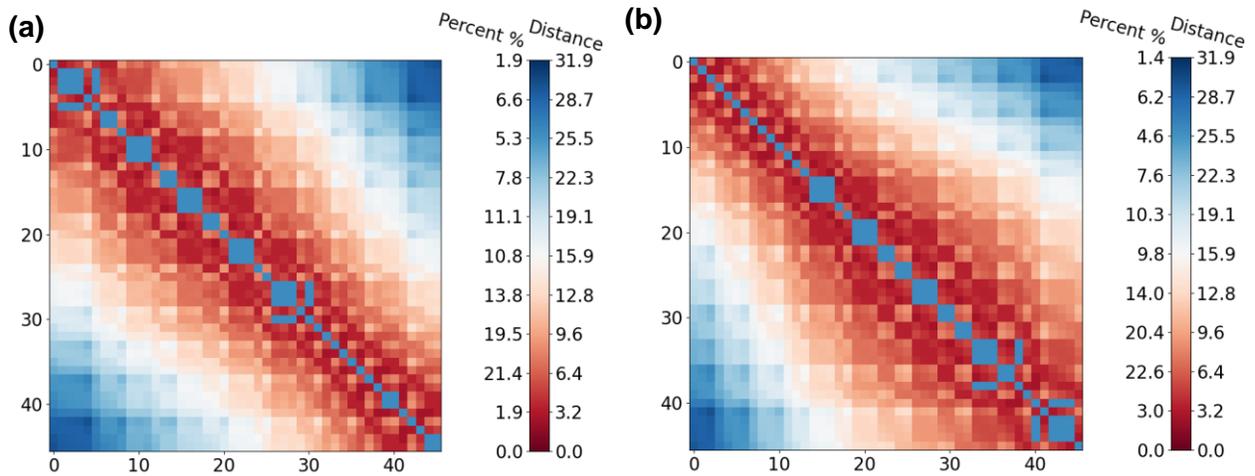

Fig. 8. The counterplot of distance of passengers from other passengers for the optimal (a) and suboptimal (b) arrangements corresponding to Figures 7a and 7b, respectively. From these plots, we see that the suboptimal arrangement has a higher red area than blue which means that the distance between passengers is higher in the optimal arrangement than the suboptimal arrangement. In the colorbar, we show the percentage that each color occupies in the plots.

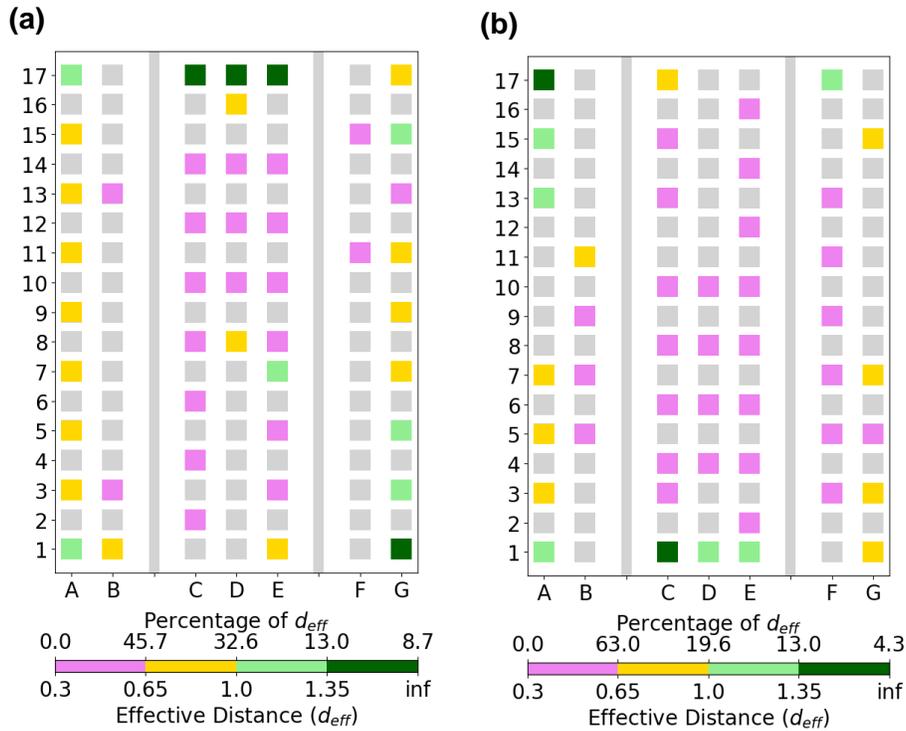

Fig. 9. The average distance of each passenger from other passengers for optimal and suboptimal arrangements corresponds to the passengers' arrangements in Figures 7a and 7b, respectively. For clarity we show the percentage that each color occupies in each plot in the corresponding colorbar. Overall, in the optimal arrangement passengers have higher effective distance from each other as compared to the suboptimal arrangement.

## Conclusion

Social distancing measures, which have been imposed by governments to reduce the health impacts of Covid-19, have had a big impact on businesses that have on-site customers as they have had to limit their on-site capacity. These businesses have mostly implemented social distancing by installing signages that block some spaces and guide customers to keep the recommended distance from other customers. However, to the best of our knowledge, they have not used modeling and optimization methods to optimize the number of people and their arrangements on their sites. In this study, we introduced a model with quantity Spread, and we proposed that by minimizing Spread, it is possible to maximize the number of people and their distances for social distancing measures on a site. To illustrate this, we studied a real-world case from the airline industry and optimized the number of passengers and the arrangements of passengers for the single and twin aisles aircraft layouts using AMC method. We showed that our model correctly arranges the family members groups close to each other and maximizes the distance between non family members. Moreover, we demonstrated that our obtained arrangements of passengers have higher distance between passengers than any suboptimal arrangement with the same number of passengers and a social distancing measure. The optimal arrangements of passengers have complex patterns, which depend on the number of groups, the number of members in each group, and the number of individual passengers. Our results clearly show that the optimal arrangements of passengers minimize the health and economic impacts of Covid-19 by maximizing the number of passengers and their distances under a social distancing measure. Moreover, we demonstrated that among the seats that are allocated to passengers, some seats have higher effective distances from other seats. These seats could be assigned to passengers who are more vulnerable to Covid-19 to reduce the health impact of Covid-19. These results clearly show that our model or its variants, along with optimization methods such as Annealing Monte Carlo method, must be utilized to minimize the health and economic impacts of Covid-19 for businesses that have on-site employees and customers. Our model could be applied to various organizations and companies with on-site people such as classrooms, theaters, restaurants, sporting events, waiting rooms, medical centers, etc.

## Methods

**Monte Carlo Simulations**. As it was discussed in the Model section our aim is to find the positions of family groups and individual passengers by minimizing Spread ($S_{tot}$) using Annealing Monte Carlo simulation. Below we use the terms individual and individuals as a replacement for person and people.

We start the simulation by assigning a random position to all individuals as the first AMC step. In each AMC step, we apply Monte Carlo simulation to find the local minimum of Spread for individuals. In every MC step, each individual (i) is allowed to minimize their own Spread by moving one step in a direction based on the Metropolis Hastings Algorithm [31]: $p = min\{1, \exp\left(-(spread_{newi} - spread_{oldi})/T\right)\}$. If the Spread of the individual cannot be accepted, the individual does not move. At each AMC step, we decrease T and change the position of an individual randomly. Then we allow all individuals to find their optimum positions based on the Metropolis-Hastings algorithm. This algorithm is "capitalistic" in nature, as each individual lowers only their own Spread. This does not automatically mean that $S_{tot}$ finds its global minimum.

The algorithm used in our code can be described in more details as below:

•Start.
Get initial individual positions randomly as trial positions.
• According to Eq. 2, calculate $S_{tot}^{old}$ of initial individual positions.
• Start AMC with high temperature ($T = 2.3 \times 10^9$, $T_1 = T$ and j = 1).
   I.    Get temperature (T), change position of an individual randomly and k = 0.
       A.  While j < number Monte Carlo steps, go to rule 0 otherwise go to rule II.

            0.     Sort individual based on maximum $spread$ and i = 1.

1. Calculate spread$_i$ from Eq. 1.
2. Take spread$_{iold}$=spread$_i$.
3. Choose the new position of individual i in its eight neighbors randomly and calculate spread$_{inew}$.
4. Check if this new position is occupied with another individual. If it is occupied before, go back to rule 3. If its eight neighbors are occupied with another individuals go to rule 10.
5. Calculate the difference between new and old spread for individual i: $\Delta_i = spread_{inew} - spread_{iold}$.
6. $p_i = \min\{1, \exp(-\Delta_i/T)\}$.
7. Get random number r.
8. If $p \geq r$, accept the new position and $spread_{iold} = spread_{inew}$.
9. If $r > p$, individual should come back to its old position.
10. i=i+1 and Sort individual based on maximum $spread$.
11. If i < number individuals, go back to rule 1 otherwise j = j+1 then go to rule A.

II. Calculate $S_{tot}^{new}$ from Eq. 2.
III. Find the difference between the old and new $S_{tot}$:
$$\Delta S = S_{tot}^{new} - S_{tot}^{old}.$$
IV. $P = \min\{1, \exp(-\Delta S/T)\}$.
V. Get random number r.
VI. If $P \geq r$, accept new trial positions and $S_{tot}^{old} = S_{tot}^{new}$.
VII. If $r > P$, go back to the old initial individual positions.
VIII. Decrease T: $T_1 = T_1 \left(0.9 + \frac{j}{300} \times 0.099\right)$ and $T = 0.026 \times \frac{T_1}{300}$.
IX. If $T > 10^{-6}$, $j = j + 1$ then go to rule I. Otherwise go to next part.

• AMC cycle gets finished.

## Acknowledgments

We thank Michel Gingras and Richard Scalettar for their valuable comments and feedback on the paper. We particularly wish to thank Michel Gingras for a critical reading of the manuscript. We also wish to thank Kate Wessels for proofreading the manuscript.

## References

[1] McKee, M., Stuckler, D. If the world fails to protect the economy, COVID-19 will damage health not just now but also in the future. *Nat Med* 26, 640–642 (2020). https://doi.org/10.1038/s41591-020-0863-y

[2] Haug, N., Geyrhofer, L., Londei, A. et al. Ranking the effectiveness of worldwide COVID-19 government interventions. Nat Hum Behav 4, 1303–1312 (2020).

[3] Anderson, R. M., Heesterbeek, H., Klinkenberg, D. & Hollingsworth, T. D. How will country-based mitigation measures influence the course of the COVID-19 epidemic? Lancet 395, 931–934 (2020).

[4] Flaxman, S. et al. Estimating the effects of non-pharmaceutical interventions on COVID-19 in Europe. Nature 584, 257–261 (2020).

[5] Opening Up America Again. https://www.whitehouse.gov/openingamerica/

[6] Pancevski B. WSJ Germany to Reopen Most of Economy in Coming Weeks as Coronavirus Recedes (2020). https://www.wsj.com/articles/germany-to-reopen-most-of-economy-in-coming-weeks-as-coronavirus-recedes-11588787063

[7] Block, P., Hoffman, M., Raabe, I.J. et al. Social network-based distancing strategies to flatten the COVID-19 curve in a post-lockdown world. Nat Hum Behav 4, 588–596 (2020).

[8] Daniel Duque, David P. Morton, Bismark Singh, Zhanwei Du, Remy Pasco, Lauren Ancel Meyers. Timing social distancing to avert unmanageable COVID-19 hospital surges. Proceedings of the National Academy of Sciences, 117 (33) 19873-19878 (2020).

[9] Gondim, J. A. M. & Machado, L. Optimal quarantine strategies for the COVID-19 pandemic in a population with a discrete age structure. Chaos Solitons Fractals 140, 110166 (2020).

[10] Lee, K., Worsnop, C. Z., Grépin, K. A. & Kamradt-Scott, A. Global coordination on cross-border travel and trade measures crucial to COVID-19 response. Lancet 395, 1593–1595 (2020).

[11] Reopening Guidance for Cleaning and Disinfecting Public Spaces, Workplaces, Businesses, Schools, and Homes. https://www.cdc.gov/coronavirus/2019-ncov/community/reopen-guidance.html

[12] Lee, J. K., & Jeong, H. W. Wearing facemasks regardless of symptoms is crucial for preventing spread of COVID-19 in hospitals. Infect. Control Hosp. Epidemiol.

[13] Lewnard, J. A., & Lo, N. C.. Scientific and ethical basis for social-distancing interventions against COVID-19. The Lancet Infectious Diseases, 20(6), 631-633 (2020).

[14] Joakim A. Weill, Matthieu Stigler, Olivier Deschenes, Michael R. Springborn. Social distancing responses to COVID-19 emergency declarations strongly differentiated by income. Proceedings of the National Academy of Science 117 (33), 19658-19660 (2020).

[15] Jay, J., Bor, J., Nsoesie, E.O. et al. Neighbourhood income and physical distancing during the COVID-19 pandemic in the United States. Nat Hum Behav 4, 1294–1302 (2020).

[16] Chinazzi, M. et al. The effect of travel restrictions on the spread of the 2019 novel coronavirus (COVID-19) outbreak. Science 368, 395–400 (2020).

[17] Wells, C. R. et al. Impact of international travel and border control measures on the global spread of the novel 2019 coronavirus outbreak. Proc. Natl Acad. Sci. USA 117, 7504–7509 (2020).

[18] Prem, K. et al. The effect of control strategies to reduce social mixing on outcomes of the COVID-19 epidemic in Wuhan, China: a modelling study. Lancet 5, e261–e270 (2020).

[19] Firth, J.A., Hellewell, J., Klepac, P. et al. Using a real-world network to model localized COVID-19 control strategies. Nat Med 26, 1616–1622 (2020).

[20] Wells, C.R., Townsend, J.P., Pandey, A. et al. Optimal COVID-19 quarantine and testing strategies. Nat Commun 12, 356 (2021). https://doi.org/10.1038/s41467-020-20742-8.

[21] Hellewell, J. et al. Feasibility of controlling COVID-19 outbreaks by isolation of cases and contacts. Lancet Glob. Health 8, E488–E496 (2020).

[22] Chang, S., Pierson, E., Koh, P.W. et al. Mobility network models of COVID-19 explain inequities and inform reopening. Nature 589, 82–87 (2021).

[23] Thunström, L., Newbold, S., Finnoff, D., Ashworth, M., & Shogren, J. The Benefits and Costs of Using Social Distancing to Flatten the Curve for COVID-19. Journal of Benefit-Cost Analysis, 11(2), 179-195 (2020).

[24] Alex Arenas, Wesley Cota, Jesús Gómez-Gardeñes, Sergio Gómez, Clara Granell, Joan T. Matamalas, David Soriano-Paños, and Benjamin Steinegger. Modeling the Spatiotemporal Epidemic Spreading of COVID-19 and the Impact of Mobility and Social Distancing Interventions, Phys. Rev. X 10, 041055 (2020).

[25] M Farboodi, G Jarosch, R Shimer. National Bureau of Economic Research Working Paper Series. Internal and external effects of social distancing in a pandemic (2020).

[26] Guan, D., Wang, D., Hallegatte, S. et al. Global supply-chain effects of COVID-19 control measures. Nat Hum Behav 4, 577–587 (2020).

[27] Kennedy, D. M., Zambrano, G. J., Wang, Y., & Neto, O. P.. Modeling the effects of intervention strategies on COVID-19 transmission dynamics. Journal of Clinical Virology, 128, 104440 (2020).

[28] Aleta, A., Martín-Corral, D., Pastore y Piontti, A. et al. Modelling the impact of testing, contact tracing and household quarantine on second waves of COVID-19. Nat Hum Behav 4, 964–971 (2020).

[29] M.E.J. Newman, T. Barkema, Monte Carlo Methods in Statistical Physics. Oxford University Press, Oxford, UK (1999).

[30] V. Granville, M. Krivanek and J. Rasson. Simulated annealing: a proof of convergence. IEEE Transactions on Pattern Analysis and Machine Intelligence, vol. 16, no. 6, pp. 652-656 (1994).

[31] C. Robert and G. Casella. The Metropois-Hastings Algorithm. Monte Carlo Statistical Methods, New York: Springer-Verlag (2001).